\documentclass[11pt]{article}
\usepackage{amsthm}
\usepackage{color}
\usepackage{amsmath,amssymb}
\title{\textbf{\large On conditional coloring of some graphs}}
\author{
P.Venkata Subba Reddy and K.Viswanathan Iyer \thanks{author for correspondence}\\
Dept. of Computer Science and Engg. \\  
National Institute of Technology\\  
Tiruchirapalli 620 015, 
India.\\
email : venkatpalagiri@gmail.com,kvi@nitt.edu} 
\date{}
\begin{document}
\maketitle
\begin{abstract}  
For integers $r$ and $k > 0\ (k>r)$, a conditional $(k, r)$-coloring of a graph $G$ is a proper $k$-coloring of $G$ such that every vertex $v$ of $G$ has at least $min\{r,d(v)\}$ differently colored neighbors, where $d(v)$ is the degree of $v$. Given an $r$, for a graph $G$ we are interested in obtaining a conditional coloring using minimum $k$ --- this value is denoted by $\chi_r(G)$, called the $r^{th}$ order conditional chromatic number. In this note, for different values of $r$ we obtain the following results: 
\begin{enumerate}
\item
For any $n,r \geq 2$, $\chi_r(G(2,n))=4$, where $G(2,n)= P_2 \  \Box \ P_n$.
\item
Let $C_n^2$ be the square of $C_n$ such that $ n \geq 3$ and $n \neq 13$,$14$ and $19$. Then 
\begin{equation*}
\chi_r(C_n^2)=
\begin{cases}
n, &  \text{if  $n \leq 5$ or both $6 \leq n \leq 9$ and $r=4$.}\\
\chi(C_n^2), &  \text{if $r=2$ .}\\
4,  &  \text{if  $n \not \equiv 3 \pmod{4}$ and $r=3$.}\\
5,  &  \text{if $n > 9$ , $n \equiv 0 \pmod{5}$ and $r=4$.}\\ 
6,    &  \text{if $n > 9$, $n \not \equiv 0 \pmod{5}$ and $r=4$}.
\end{cases}
\end{equation*}
\item
For $n,m \geq 2$, let $P_n \otimes P_m$ be the strong product of graphs $P_n$ and $P_m$. Then 
\begin{equation*}
\chi_r(P_n \otimes P_m)=
\begin{cases}
4, &  \text{if  $r \leq 3$.}\\
\Delta+1, &  \text{if $r=\Delta$ .}
\end{cases} 
\end{equation*}
\item
Let $W(t,n)$ be a $(t,n)$-web graph, for $t \geq 1$ and $n \geq 3$. $W(t,n)$    consists of $t$ induced cycles $C_n$ such that no two $C_n$'s have a vertex in common and is constructed recursively thus. Let $W(1,n)=W_{n+1}$, the wheel graph on $n+1$ vertices. We denote by $v_{0,0}$ the vertex of $W(1,n)$ with degree $n$. Assume that $W(t,n)$ has been obtained. Let $v_{t,1},v_{t,2},\ldots,v_{t,n}$ be the induced cycle in $W(t,n)$ such that for all $1 \leq i \leq n,\; d(v_{t,i})=3$ and  $v_{t,i}$ is adjacent to $v_{t,j}$ only if $|i-j|=1 \; \text{or} \; n-1$. We construct $W(t+1,n)$ from $W(t,n)$ with the following vertex and edge sets: 
$V(W(t+1,n)) =V(W(t,n)) \cup \{v_{t+1,1},v_{t+1,2},\ldots,v_{t+1,n}\}$ and 
$E(W(t+1,n)) =E(W(t,n)) \cup \{(v_{t,i},v_{t+1,i}): 1 \leq i \leq n \} \cup E',$ 
where $E'=\{(v_{t+1,i},v_{t+1,j}): 1 \leq i,j \leq n \; \text{and} \; j-i=1 \; \text{or} \; n-1\}$.\\
Then $\chi_2(W(t,n))=4$.
\end{enumerate} 
\end{abstract} 
\section{Introduction}
Let $G= (V(G),E(G))$ be a simple, connected, undirected graph. $\Delta(G)$ and $\omega(G)$ (or simply $\Delta$ and $\omega$) denote, respectively the  maximum degree and the clique number of a graph $G$. For a vertex $v \in V(G)$, the (\textit{open}) \textit{neighborhood} of $v$ in $G$ is $N(v)$= \{$u \in V(G):(u,v) \in E(G)$\}, the \textit{closed neighborhood} of $v$ is defined as $N[v]=N(v) \cup \{v\}$ and the degree of $v$ is $d(v)=|N(v)|$.  For an integer $k>0$, a \textit{proper} $k$-coloring of a graph $G$ is a surjective mapping $c \colon V(G) \to \{1,\ldots,k \}$ such that if $(u,v) \in E(G)$, then $c(u) \neq c(v)$. The smallest $k$ such that $G$ has a proper $k$-coloring is the \textit{chromatic number} of $G$, denoted by $\chi(G)$. For integers $k>0$ and $r>0$, a \textit{conditional} $(k,r)$-coloring of a graph $G$ is a surjective mapping $c \colon V(G) \to \{1,\ldots,k \}$ satisfying the following conditions:
\begin{quote}
(C1) if $(u,v) \in E(G)$, then $c(u) \neq c(v)$; \\
(C2) for any $v \in V(G)$, $|c(N(v))| \geq $ min \{$r,d(v)$ \}, where (and in what follows) $c(S)= \{c(u) : u \in S$ for a set $S \subseteq V(G)$\}. 
\end{quote}
For a given integer $r>0$, the smallest integer $k$ such that $G$ has a conditional $(k,r)$-coloring is called the \textit{rth-order conditional chromatic number} of $G$, denoted by $\chi_r(G)$. In particular when $r=2,$ it is called \textit{dynamic chromatic number} denoted by $\chi_d(G)$. In ~\cite{Li2}, the authors proved that for $r\le 2 < k$ conditional $(k, r)$-coloring of a graph is NP-complete. By the definition of $\chi_r(G)$, it is clear that $\chi(G)=\chi_1(G)$ and so conditional coloring is a generalization of traditional graph coloring. However unlike in  traditional coloring, if $G'$ is a subgraph of $G$ then $\chi_r(G') \leq \chi_r(G)$ need not be true. Also, it can be easily seen that $\chi_{r'}(G') \leq \chi_\Delta(G)$ for any positive integer $r'$. By $P_n,C_n$ and $K_n$ we denote a path, a cycle and a complete graph respectively of $n$ vertices. For undefined terms and notations see standard texts in graph theory such as ~\cite{Bondy,Gary,Gol}.
\section{Some Results on Conditional Colorability of Graphs}
\newtheorem{lem1}{Lemma}
\begin{lem1}
Let $G$ be a graph which has two adjacent vertices $u$ and $v$ such that $d(u),d(v) \leq r \leq \Delta; \forall u' \in N(u) \setminus \{v\},N(v) \setminus \{u\} \subset N(u')$ and for all $v' \in N(v) \setminus \{u\},N(u) \setminus \{v\} \subset N(v')$. Then $\chi_r(G) \geq d(u)+d(v)$.
\end{lem1}
\begin{proof}
The proof by contradiction is straightforward.
 \end{proof}
\newtheorem{prop1}{Proposition}
\begin{prop1}
For any $n,r \geq 2$, $\chi_r(G(2,n))=4$, where $G(2,n)= P_2 \  \Box \ P_n$.
\end{prop1}
\begin{proof}
The result is trivial for $n=2$. Since $G(2,n)$ has two adjacent vertices with degree two each satisfying conditions in Lemma $1$ we have $\chi_r(G(2,n)) \geq 4$. Moreover, it is possible to find a conditional $(4,r)$-coloring of $G(2,n)$ as can be seen in Fig. 1. Thus $\chi_r(G(2,n)) \leq 4$, and altogether we have $\chi_r(G(2,n))= 4$.  
\end{proof}
\setlength{\unitlength}{0.5mm}
\begin{center}
\begin{picture}(200, 40)
\put(15,10){\circle*{2}}
\put(35,10){\circle*{2}}
\put(55,10){\circle*{2}}
\put(75,10){\circle*{2}}
\put(95,10){\circle*{2}}
\put(115,10){\circle*{2}}
\put(135,10){\circle*{2}}
\put(155,10){\circle*{2}}
\put(175,10){\circle*{2}}
\put(195,10){\circle*{2}}
\put(15,10){\line(1,0){20}}
\put(35,10){\line(1,0){20}}
\put(55,10){\line(1,0){20}}
\put(75,10){\line(1,0){20}}
\put(95,10){\line(1,0){20}}
\put(115,10){\line(1,0){20}}
\put(135,10){\line(1,0){20}}
\put(155,10){\line(1,0){20}}
\put(175,10){\line(1,0){20}}
\put(15,30){\circle*{2}}
\put(35,30){\circle*{2}}
\put(55,30){\circle*{2}}
\put(75,30){\circle*{2}}
\put(95,30){\circle*{2}}
\put(115,30){\circle*{2}}
\put(135,30){\circle*{2}}
\put(155,30){\circle*{2}}
\put(175,30){\circle*{2}}
\put(195,30){\circle*{2}}
\put(15,30){\line(1,0){20}}
\put(35,30){\line(1,0){20}}
\put(55,30){\line(1,0){20}}
\put(75,30){\line(1,0){20}}
\put(95,30){\line(1,0){20}}
\put(115,30){\line(1,0){20}}
\put(135,30){\line(1,0){20}}
\put(155,30){\line(1,0){20}}
\put(175,30){\line(1,0){20}}
\put(15,10){\line(0,1){20}}
\put(35,10){\line(0,1){20}}
\put(55,10){\line(0,1){20}}
\put(75,10){\line(0,1){20}}
\put(95,10){\line(0,1){20}}
\put(115,10){\line(0,1){20}}
\put(135,10){\line(0,1){20}}
\put(155,10){\line(0,1){20}}
\put(175,10){\line(0,1){20}}
\put(195,10){\line(0,1){20}}
\put(14,34){\scriptsize{\textbf{1}}}
\put(34,34){\scriptsize{\textbf{2}}}
\put(54,34){\scriptsize{\textbf{3}}}
\put(74,34){\scriptsize{\textbf{4}}}
\put(94,34){\scriptsize{\textbf{1}}}
\put(114,34){\scriptsize{\textbf{2}}}
\put(134,34){\scriptsize{\textbf{3}}}
\put(154,34){\scriptsize{\textbf{4}}}
\put(174,34){\scriptsize{\textbf{1}}}
\put(194,34){\scriptsize{\textbf{2}}}
\put(14,2){\scriptsize{\textbf{3}}}
\put(34,2){\scriptsize{\textbf{4}}}
\put(54,2){\scriptsize{\textbf{1}}}
\put(74,2){\scriptsize{\textbf{2}}}
\put(94,2){\scriptsize{\textbf{3}}}
\put(114,2){\scriptsize{\textbf{4}}}
\put(134,2){\scriptsize{\textbf{1}}}
\put(154,2){\scriptsize{\textbf{2}}} 
\put(174,2){\scriptsize{\textbf{3}}}
\put(194,2){\scriptsize{\textbf{4}}}
\put(196,30){\circle*{0.1}}
\put(198,30){\circle*{0.1}}
\put(200,30){\circle*{0.1}}
\put(202,30){\circle*{0.1}}
\put(204,30){\circle*{0.1}}
\put(206,30){\circle*{0.1}}
\put(208,30){\circle*{0.1}}
\put(210,30){\circle*{0.1}}
\put(196,10){\circle*{0.1}}
\put(198,10){\circle*{0.1}} 
\put(200,10){\circle*{0.1}}
\put(202,10){\circle*{0.1}}
\put(204,10){\circle*{0.1}}
\put(206,10){\circle*{0.1}}
\put(208,10){\circle*{0.1}}
\put(210,10){\circle*{0.1}}
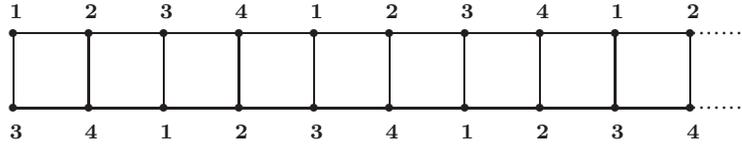
\put(28,-10){\footnotesize{\textbf{Fig.1.}} \footnotesize{Conditional $(4,r)$-coloring of $G(2,n)$}}
\end{picture}
\end{center} 

\subsection{Computing $\chi_r(C_n^2)$}
The \textit{distance} $d(u,v)$ between two vertices $u$ and $v$ is the minimum length of a path between $u$ and $v$. The \textit{square} of a graph $G$, denoted by $G^2$, has the same vertex set as $G$ with the edge set $E(G^2)=\{(u,v): d(u,v) \leq 2 \; \text{in} \; G\}$. In particular $C_n^2$ denotes the square of the cycle $C_n$. In this section we determine $\chi_r(C_n^2)$ for different values of $r$. The \textit{diameter} of a graph $G$ is $diam(G)=max \{d(u,v):u,v \in V(G) $\}. We assume that the vertices of $C_n^2$ are labelled as $v_1, \ldots , v_n$ such that for all $3 \leq i \leq n-2, \; N(v_i)=\{v_{i-2},v_{i-1},v_{i+1},v_{i+2}\}$; $N(v_1)=\{v_{n},v_{n-1},2,3\}, \; N(v_2)=\{v_{n},1,3,4\}, \; N(v_{n-1})=\{v_{n-3},v_{n-2},n,1\}$ and $N(v_n)=\{1,2,$  $v_{n-2},v_{n-1}\}$. Clearly $\Delta(C_n^2)=4$. 
\newtheorem{thm1}{Theorem}
\begin{thm1}
Let $C_n^2$ be the square of $C_n$ such that $ n \geq 3$ and $n \neq 13$,$14$ and $19$. Then 
\begin{equation*}
\chi_r(C_n^2)=
\begin{cases}
n, &  \text{if  $n \leq 5$ or both $6 \leq n \leq 9$ and $r=4$.}\\
\chi(C_n^2), &  \text{if $r=2$ .}\\
4,  &  \text{if  $n \not \equiv 3 \pmod{4}$ and $r=3$.}\\
5,  &  \text{if $n > 9$ , $n \equiv 0 \pmod{5}$ and $r=4$.}\\ 
6,    &  \text{if $n > 9$, $n \not \equiv 0 \pmod{5}$ and $r=4$}.
\end{cases}
\end{equation*}
\end{thm1} 
\begin{proof} 
From ~\cite{Lai1} we know that $\chi_{r'}(G) \geq\,$ min $\{r',\Delta\}+1$. Taking $G=C_n^2$ and $r'=r$ we have $\chi_r(C_n^2) \geq $ min $\{r,4\}+1$. We now consider the different cases below. \\
\textbf{Case 1:} $n \leq 5$. In this case $C_n^2 \cong K_n$. Since $\chi(C_n^2) \leq \chi_r(C_n^2) \leq |V(C_n^2)|$ and $\chi(K_n)=n$; hence the result.\\
\textbf{Case 2:} $r=2$. It can be easily seen that every vertex of $C_n^2$ is contained in $K_3$. So in any proper $k$-coloring $c$ of $C_n^2$, every vertex $v$ has at least two distinctly colored neighbors. Hence conditional $(\chi(C_n^2),2)$-coloring of $C_n^2$ exists, so $\chi_r(C_n^2) \leq \chi(C_n^2)$. But from ~\cite{Lai1} we know that $\chi(G) \leq \chi_r(G)$, hence we have $\chi_r(C_n^2) \geq \chi(C_n^2)$. Therefore $\chi_r(C_n^2) = \chi(C_n^2)$. Also, one can easily see that $\chi(C_n^2)=3$, if $n \equiv 0 \pmod{3}$ and $4$ otherwise. \\
\textbf{Case 3:} $n \not \equiv 3 \pmod{4}$ and $r=3$. Clearly $\chi_r(C_n^2) \geq $ min $\{r,4\}+1=4$.  We now define the coloring assignment $c \colon V(C_n^2) \to \{1,2,3,4 \}$ as follows: \\
\hspace*{2cm}if  $n \equiv 0 \pmod{4}$ then 
$$c(v_i)=(i-1) \mod 4 +1.$$
\hspace*{2cm}if  $n \equiv 1 \pmod{4}$ then redefine $c$ at $v_{n-4},\ldots,v_n$ as
\begin{equation*}
c(v_i)=
\begin{cases}  
1, &  \text{if $i =n-3$.}\\
2, &  \text{if $i =n-4$ or $n-1$.}\\
3, &  \text{if $i =n-2$ .}\\  
4, &  \text{if $i =n$ .}
\end{cases}
\end{equation*}
\hspace*{2cm}if $n \equiv 2 \pmod{4}$ then redefine $c$ at $v_{n-6},\ldots,v_n$ as
\begin{equation*}
c(v_i)=
\begin{cases} 
1, &  \text{if $i =n-6$ or $n-3$.}\\
2, &  \text{if $i =n-4$ or $n-1$.}\\ 
3, &  \text{if $i =n-2$ .}\\
4, &  \text{if $i =n-5$ or $n$.}
\end{cases}
\end{equation*}
From the above assignment it can be checked that $c$ satisfies (C1). Clearly for all $1 \leq i \leq n$, $c(N(v_i))=\{1,2,3,4\} \setminus \{c(v_i)\}$, hence (C2) is also satisfied. Therefore $c$ defines a conditional $(4,r)$-coloring of $C_n^2$. Hence $\chi_3(C_n^2) \leq  4$ and altogether, we have $\chi_r(C_n^2) = 4$.\\
\textbf{Case 4:} $6 \leq n \leq 9$ and $r=4$. Assume the contradiction and let $\chi_r(C_n^2) =n-1$. Also let $c \colon V(C_n^2) \to \{1,\ldots,n-1 \}$ be a possible conditional $(n-1,r)$- coloring of $C_n^2$. Since $diam(C_n^2) =2$, there exist three vertices $u,v,w \in V(C_n^2)$ such that $c(u)=c(v)$ and $u,v \in N(w)$. This implies $|c(N(w))|<$ min $ \{r,d(w)\}$, so (C2) is violated at $w$. Hence $\chi_r(C_n^2) \geq n$. But we know that $\chi_r(C_n^2) \leq |V(C_n^2)|=n$; hence the result. \\
\textbf{Case 5:} $n > 9$, $n \equiv 0 \pmod{5}$ and $r=4$. Clearly $\chi_r(C_n^2) \geq $ min $ \{r,4\}+1=5$.  We now define the coloring assignment $c \colon V(C_n^2) \to \{1,2,3,4,5 \}$ as follows: 
$$c(v_i)=(i-1) \mod 5 +1.$$
In the above assignment for all $1 \leq i \leq n$, $c(N[v_i])=\{1,2,3,4,5\}$; this implies  $|c(N[v_i])|= |N[v_i]|$. Hence $c$ satisfies both (C1) and (C2). So $c$ defines a conditional $(5,r)$-coloring of $C_n^2$. Hence $\chi_r(C_n^2) \leq  5$ and altogether we have $\chi_r(C_n^2) = 5$.\\
\textbf{Case 6:} $n > 9$, $n \not \equiv 0 \pmod{5}$ and $r=4$. Let $k=\chi_r(C_n^2)$. We assume that $k=5$. Let $c \colon V((C_n^2)) \to \{1,2,3,4,5 \}$ be a possible conditional $(5,r)$-coloring of $C_n^2$. In the circular order $v_1,\ldots,v_n$, every five consecutive vertices must be colored differently, otherwise (C1) or (C2) or both are violated at the center vertex. So w.l.o.g., we may assume that $c(v_i)=(i-1) \mod 5 +1$ for all $1 \leq i \leq n$. If $n \equiv 1 \pmod{5}$ then we would have $c(v_{1})=c(v_{n})$, contrary to (C1); If $n \equiv 2 \pmod{5}$ then we would have $c(v_{1})=c(v_{n-1})$ and $c(v_{2})=c(v_{n})$, a violation of (C1); If $n \equiv 3 \pmod{5}$ then we would have $c(v_{1})=c(v_{n-2})$,\;$c(v_{2})=c(v_{n-1})$  and $c(v_{3})=c(v_{n})$, a violation of (C2) at $v_{1},v_{2},v_{n-1}$ and  $v_{n}$; If $n \equiv 4 \pmod{5}$ then we would have $c(v_{1})=c(v_{n-3})$ and $c(v_{2})=c(v_{n-2})$, a violation of (C2) at $v_{1},v_{2},v_{n-1}$ and  $v_{n}$. Therefore, we must have $k \geq 6$. To show that $k=6$, it suffices to construct a conditional $(6,r)$-coloring of $C_n^2$.\\
We now define the coloring assignment $c \colon V(C_n^2) \to \{1,\ldots,6 \}$ for admissible $i$ as follows: 
\begin{equation*}
c(v_i)=
\begin{cases} 
(i-1) \mod 5 +1, &  \text{if $1 \leq i \leq l-1$.}\\
(i-l) \mod 6 +1, &  \text{if $l \leq i \leq n$.} 
\end{cases}
\end{equation*}
where, $\; l=n+1-6(n \mod 5)$.  \\
Similar to case 5, in the above assignment it can be easily verified that for all $1 \leq i \leq n$, $|c(N[v_i])|= |N[v_i]|$. Hence $c$ satisfies both (C1) and (C2). So $c$ defines a conditional $(6,r)$-coloring of $C_n^2$. Hence $\chi_r(C_n^2) \leq  6$ and altogether we have $\chi_r(C_n^2) = 6 $.
\end{proof} 
 

\subsection{Computation of $\chi_r(P_n \otimes P_m)$}
The \textit{strong product} $G \otimes H$ of graphs $G$ and $H$ is defined as follows: $V(G \otimes H)=$ Cartesian product of $V(G)$ and $V(H)$ and $E(G \otimes H)=E_c \cup E_d$, where $E_c=\{((x_1,x_2),(y_1,y_2)): (x_1,y_1) \in E(G)\; \text{and} \; x_2=y_2, \; \text{\textit{or}} \; (x_2,y_2) \in E(H) \; \text{and} \; x_1=y_1  \}$ and 
$E_d=\{((x_1,x_2),(y_1,y_2)): (x_1,y_1) \in E(G) \; \text{and} \; (x_2,y_2) \in E(H)\}$.
\newtheorem{thm2}[thm1]{Theorem}
\begin{thm2}
For $n,m \geq 2$, let $P_n \otimes P_m$ be the strong product of graphs $P_n$ and $P_m$. Then 
\begin{equation*}
\chi_r(P_n \otimes P_m)=
\begin{cases}
4, &  \text{if  $r \leq 3$.}\\
\Delta+1, &  \text{if $r=\Delta$ .}
\end{cases} 
\end{equation*}
\end{thm2}
\begin{proof} 
Since $P_n \otimes P_m=P_m \otimes P_n$ (strong product of graphs is symmetric), we assume $n \leq m$ w.l.o.g. We will identify each vertex $v$ of $P_n \otimes P_m$ by its coordinates $(x_i,y_i),0 \leq x_i \leq n-1$ and $0 \leq y_i \leq n-1$.\\ 
\textbf{Case 1:}  $r \leq 3$. It is easy to see that every vertex of $P_n \otimes P_m$ is contained in $K_4$. So $\chi_r(P_n \otimes P_m) \geq 4$. We now define the coloring assignment $c \colon V(P_n \otimes P_m) \to \{1,2,3,4 \}$ as follows: 
$$c(v=(x_i,y_i))=x_i \mod 2 + 2(y_i \mod 2)+1.$$
It can be easily checked that $c$ defines a conditional $(4,r)$-coloring of $P_n \otimes P_m$. Therefore $\chi(P_n \otimes P_m) \leq \chi_r(P_n \otimes P_m) \leq 4$, and altogether $\chi_r(P_n \otimes P_m) = \chi(P_n \otimes P_m)=4$.\\
\textbf{Case 2:}  $r = \Delta$. The result is trivial for $n=2$ and $m=2$. 
From ~\cite{Lai1} we know that $\chi_{r}(G) \geq\,$ min $\{r,\Delta\}+1$. Taking $G=P_n \otimes P_m $ and $r=\Delta$ we have $\chi_\Delta(P_n \otimes P_m) \geq \Delta+1$. Now let us show that $\chi_\Delta(P_n \otimes P_m) \leq \Delta+1$.  We now define the coloring assignment $c \colon V(P_n \otimes P_m) \to \{1,\ldots,\Delta+1 \}$ as follows: 
$$c(v=(x_i,y_i))=x_i \mod 3 + 3(y_i \mod 3)+1.$$
If $n=2$ and $m \neq n$ then $\Delta(P_n \otimes P_m)=5$. Fig. 2 shows an example of this coloring using $6$ colors for $n=2$ and $m=11$. If $n \geq 3$  then $\Delta(P_n \otimes P_m)=8$ and  Fig. 3 shows an example of this coloring with $9$ colors for $n=5$ and $m=11$. From the Fig. 2 and Fig. 3 it is evident that $c$ satisfies both (C1) and (C2), hence defines a conditional $(\Delta+1,\Delta)$-coloring of $P_n \otimes P_m$. Therefore $\chi_\Delta(P_n \otimes P_m) \leq \Delta+1$, and altogether $\chi_\Delta(P_n \otimes P_m) = \Delta+1$.
\end{proof} 
\setlength{\unitlength}{0.5mm}
\begin{center}
\begin{picture}(220, 40)
\put(15,10){\circle*{2}}
\put(35,10){\circle*{2}}
\put(55,10){\circle*{2}}
\put(75,10){\circle*{2}}
\put(95,10){\circle*{2}}
\put(115,10){\circle*{2}}
\put(135,10){\circle*{2}}
\put(155,10){\circle*{2}}
\put(175,10){\circle*{2}}
\put(195,10){\circle*{2}}
\put(215,10){\circle*{2}}
\put(15,10){\line(1,0){20}}
\put(35,10){\line(1,0){20}}
\put(55,10){\line(1,0){20}}
\put(75,10){\line(1,0){20}}
\put(95,10){\line(1,0){20}}
\put(115,10){\line(1,0){20}}
\put(135,10){\line(1,0){20}}
\put(155,10){\line(1,0){20}}
\put(175,10){\line(1,0){20}}
\put(195,10){\line(1,0){20}}
\put(15,30){\circle*{2}}
\put(35,30){\circle*{2}}
\put(55,30){\circle*{2}}
\put(75,30){\circle*{2}}
\put(95,30){\circle*{2}}
\put(115,30){\circle*{2}}
\put(135,30){\circle*{2}}
\put(155,30){\circle*{2}}
\put(175,30){\circle*{2}}
\put(195,30){\circle*{2}}
\put(215,30){\circle*{2}}
\put(15,30){\line(1,0){20}}
\put(35,30){\line(1,0){20}}
\put(55,30){\line(1,0){20}}
\put(75,30){\line(1,0){20}}
\put(95,30){\line(1,0){20}}
\put(115,30){\line(1,0){20}}
\put(135,30){\line(1,0){20}}
\put(155,30){\line(1,0){20}}
\put(175,30){\line(1,0){20}}
\put(195,30){\line(1,0){20}}
\put(15,10){\line(0,1){20}}
\put(35,10){\line(0,1){20}}
\put(55,10){\line(0,1){20}}
\put(75,10){\line(0,1){20}}
\put(95,10){\line(0,1){20}}
\put(115,10){\line(0,1){20}}
\put(135,10){\line(0,1){20}}
\put(155,10){\line(0,1){20}}
\put(175,10){\line(0,1){20}}
\put(195,10){\line(0,1){20}}
\put(215,10){\line(0,1){20}}
\put(14,34){\scriptsize{4}}
\put(34,34){\scriptsize{5}}
\put(54,34){\scriptsize{6}}
\put(74,34){\scriptsize{4}}
\put(94,34){\scriptsize{5}}
\put(114,34){\scriptsize{6}}
\put(134,34){\scriptsize{4}}
\put(154,34){\scriptsize{5}}
\put(174,34){\scriptsize{6}}
\put(194,34){\scriptsize{4}}
\put(214,34){\scriptsize{5}}
\put(14,2){\scriptsize{1}}
\put(34,2){\scriptsize{2}}
\put(54,2){\scriptsize{3}}
\put(74,2){\scriptsize{1}}
\put(94,2){\scriptsize{2}}
\put(114,2){\scriptsize{3}}
\put(134,2){\scriptsize{1}}
\put(154,2){\scriptsize{2}} 
\put(174,2){\scriptsize{3}}
\put(194,2){\scriptsize{1}}
\put(214,2){\scriptsize{2}}
\put(15,10){\line(1,1){20}}
\put(35,10){\line(1,1){20}}
\put(55,10){\line(1,1){20}}
\put(75,10){\line(1,1){20}}
\put(95,10){\line(1,1){20}}
\put(115,10){\line(1,1){20}}
\put(135,10){\line(1,1){20}}
\put(155,10){\line(1,1){20}}
\put(175,10){\line(1,1){20}}
\put(195,10){\line(1,1){20}}
\put(35,10){\line(-1,1){20}}
\put(55,10){\line(-1,1){20}}
\put(75,10){\line(-1,1){20}}
\put(95,10){\line(-1,1){20}}
\put(115,10){\line(-1,1){20}}
\put(135,10){\line(-1,1){20}}
\put(155,10){\line(-1,1){20}}
\put(175,10){\line(-1,1){20}}
\put(195,10){\line(-1,1){20}}
\put(215,10){\line(-1,1){20}}
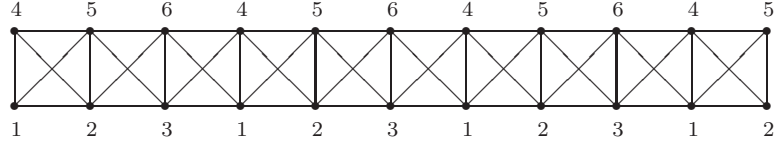
\put(28,-10){\footnotesize{\textbf{Fig.2.}} \footnotesize{Conditional $(6,5)$-coloring of $P_2 \otimes P_{11}$}}
\end{picture}
\end{center} 
\vspace{8mm}
\setlength{\unitlength}{0.5mm}
\begin{center}
\begin{picture}(220,90)
\put(15,10){\circle*{2}}
\put(35,10){\circle*{2}}
\put(55,10){\circle*{2}}
\put(75,10){\circle*{2}}
\put(95,10){\circle*{2}}
\put(115,10){\circle*{2}}
\put(135,10){\circle*{2}}
\put(155,10){\circle*{2}}
\put(175,10){\circle*{2}}
\put(195,10){\circle*{2}}
\put(215,10){\circle*{2}}
\put(15,30){\circle*{2}}
\put(35,30){\circle*{2}}
\put(55,30){\circle*{2}}
\put(75,30){\circle*{2}}
\put(95,30){\circle*{2}}
\put(115,30){\circle*{2}}
\put(135,30){\circle*{2}}
\put(155,30){\circle*{2}}
\put(175,30){\circle*{2}}
\put(195,30){\circle*{2}}
\put(215,30){\circle*{2}}
\put(15,50){\circle*{2}}
\put(35,50){\circle*{2}}
\put(55,50){\circle*{2}}
\put(75,50){\circle*{2}}
\put(95,50){\circle*{2}}
\put(115,50){\circle*{2}}
\put(135,50){\circle*{2}}
\put(155,50){\circle*{2}}
\put(175,50){\circle*{2}}
\put(195,50){\circle*{2}}
\put(215,50){\circle*{2}}
\put(15,70){\circle*{2}}
\put(35,70){\circle*{2}}
\put(55,70){\circle*{2}}
\put(75,70){\circle*{2}}
\put(95,70){\circle*{2}}
\put(115,70){\circle*{2}}
\put(135,70){\circle*{2}}
\put(155,70){\circle*{2}}
\put(175,70){\circle*{2}}
\put(195,70){\circle*{2}}
\put(215,70){\circle*{2}}
\put(15,90){\circle*{2}}
\put(35,90){\circle*{2}}
\put(55,90){\circle*{2}}
\put(75,90){\circle*{2}}
\put(95,90){\circle*{2}}
\put(115,90){\circle*{2}}
\put(135,90){\circle*{2}}
\put(155,90){\circle*{2}}
\put(175,90){\circle*{2}}
\put(195,90){\circle*{2}}
\put(215,90){\circle*{2}}
\put(15,10){\line(1,0){20}}
\put(35,10){\line(1,0){20}}
\put(55,10){\line(1,0){20}}
\put(75,10){\line(1,0){20}}
\put(95,10){\line(1,0){20}}
\put(115,10){\line(1,0){20}}
\put(135,10){\line(1,0){20}}
\put(155,10){\line(1,0){20}}
\put(175,10){\line(1,0){20}}
\put(195,10){\line(1,0){20}}
\put(15,30){\line(1,0){20}}
\put(35,30){\line(1,0){20}}
\put(55,30){\line(1,0){20}}
\put(75,30){\line(1,0){20}}
\put(95,30){\line(1,0){20}}
\put(115,30){\line(1,0){20}}
\put(135,30){\line(1,0){20}}
\put(155,30){\line(1,0){20}}
\put(175,30){\line(1,0){20}}
\put(195,30){\line(1,0){20}}
\put(15,50){\line(1,0){20}}
\put(35,50){\line(1,0){20}}
\put(55,50){\line(1,0){20}}
\put(75,50){\line(1,0){20}}
\put(95,50){\line(1,0){20}}
\put(115,50){\line(1,0){20}}
\put(135,50){\line(1,0){20}}
\put(155,50){\line(1,0){20}}
\put(175,50){\line(1,0){20}}
\put(195,50){\line(1,0){20}}
\put(15,70){\line(1,0){20}}
\put(35,70){\line(1,0){20}}
\put(55,70){\line(1,0){20}}
\put(75,70){\line(1,0){20}}
\put(95,70){\line(1,0){20}}
\put(115,70){\line(1,0){20}}
\put(135,70){\line(1,0){20}}
\put(155,70){\line(1,0){20}}
\put(175,70){\line(1,0){20}}
\put(195,70){\line(1,0){20}}
\put(15,90){\line(1,0){20}}
\put(35,90){\line(1,0){20}}
\put(55,90){\line(1,0){20}}
\put(75,90){\line(1,0){20}}
\put(95,90){\line(1,0){20}}
\put(115,90){\line(1,0){20}}
\put(135,90){\line(1,0){20}}
\put(155,90){\line(1,0){20}}
\put(175,90){\line(1,0){20}}
\put(195,90){\line(1,0){20}}
\put(15,10){\line(0,1){20}}
\put(35,10){\line(0,1){20}}
\put(55,10){\line(0,1){20}}
\put(75,10){\line(0,1){20}}
\put(95,10){\line(0,1){20}}
\put(115,10){\line(0,1){20}}
\put(135,10){\line(0,1){20}}
\put(155,10){\line(0,1){20}}
\put(175,10){\line(0,1){20}}
\put(195,10){\line(0,1){20}}
\put(215,10){\line(0,1){20}}
\put(15,30){\line(0,1){20}}
\put(35,30){\line(0,1){20}}
\put(55,30){\line(0,1){20}}
\put(75,30){\line(0,1){20}}
\put(95,30){\line(0,1){20}}
\put(115,30){\line(0,1){20}}
\put(135,30){\line(0,1){20}}
\put(155,30){\line(0,1){20}}
\put(175,30){\line(0,1){20}}
\put(195,30){\line(0,1){20}}
\put(215,30){\line(0,1){20}}
\put(15,50){\line(0,1){20}}
\put(35,50){\line(0,1){20}}
\put(55,50){\line(0,1){20}}
\put(75,50){\line(0,1){20}}
\put(95,50){\line(0,1){20}}
\put(115,50){\line(0,1){20}}
\put(135,50){\line(0,1){20}}
\put(155,50){\line(0,1){20}}
\put(175,50){\line(0,1){20}}
\put(195,50){\line(0,1){20}}
\put(215,50){\line(0,1){20}}
\put(15,70){\line(0,1){20}}
\put(35,70){\line(0,1){20}}
\put(55,70){\line(0,1){20}}
\put(75,70){\line(0,1){20}}
\put(95,70){\line(0,1){20}}
\put(115,70){\line(0,1){20}}
\put(135,70){\line(0,1){20}}
\put(155,70){\line(0,1){20}}
\put(175,70){\line(0,1){20}}
\put(195,70){\line(0,1){20}}
\put(215,70){\line(0,1){20}}
\put(15,10){\line(1,1){20}}
\put(35,10){\line(1,1){20}}
\put(55,10){\line(1,1){20}}
\put(75,10){\line(1,1){20}}
\put(95,10){\line(1,1){20}}
\put(115,10){\line(1,1){20}}
\put(135,10){\line(1,1){20}}
\put(155,10){\line(1,1){20}}
\put(175,10){\line(1,1){20}}
\put(195,10){\line(1,1){20}}
\put(15,30){\line(1,1){20}}
\put(35,30){\line(1,1){20}}
\put(55,30){\line(1,1){20}}
\put(75,30){\line(1,1){20}} 
\put(95,30){\line(1,1){20}}
\put(115,30){\line(1,1){20}}
\put(135,30){\line(1,1){20}}
\put(155,30){\line(1,1){20}}
\put(175,30){\line(1,1){20}}
\put(195,30){\line(1,1){20}}
\put(15,50){\line(1,1){20}}
\put(35,50){\line(1,1){20}}
\put(55,50){\line(1,1){20}}
\put(75,50){\line(1,1){20}}
\put(95,50){\line(1,1){20}}
\put(115,50){\line(1,1){20}}
\put(135,50){\line(1,1){20}}
\put(155,50){\line(1,1){20}}
\put(175,50){\line(1,1){20}}
\put(195,50){\line(1,1){20}}
\put(15,70){\line(1,1){20}}
\put(35,70){\line(1,1){20}}
\put(55,70){\line(1,1){20}}
\put(75,70){\line(1,1){20}}
\put(95,70){\line(1,1){20}}
\put(115,70){\line(1,1){20}}
\put(135,70){\line(1,1){20}}
\put(155,70){\line(1,1){20}}
\put(175,70){\line(1,1){20}}
\put(195,70){\line(1,1){20}}
\put(35,10){\line(-1,1){20}}
\put(55,10){\line(-1,1){20}}
\put(75,10){\line(-1,1){20}}
\put(95,10){\line(-1,1){20}}
\put(115,10){\line(-1,1){20}}
\put(135,10){\line(-1,1){20}}
\put(155,10){\line(-1,1){20}}
\put(175,10){\line(-1,1){20}}
\put(195,10){\line(-1,1){20}}
\put(215,10){\line(-1,1){20}}
\put(35,30){\line(-1,1){20}}
\put(55,30){\line(-1,1){20}}
\put(75,30){\line(-1,1){20}}
\put(95,30){\line(-1,1){20}}
\put(115,30){\line(-1,1){20}}
\put(135,30){\line(-1,1){20}}
\put(155,30){\line(-1,1){20}}
\put(175,30){\line(-1,1){20}}
\put(195,30){\line(-1,1){20}}
\put(215,30){\line(-1,1){20}}
\put(35,50){\line(-1,1){20}}
\put(55,50){\line(-1,1){20}}
\put(75,50){\line(-1,1){20}}
\put(95,50){\line(-1,1){20}}
\put(115,50){\line(-1,1){20}}
\put(135,50){\line(-1,1){20}}
\put(155,50){\line(-1,1){20}}
\put(175,50){\line(-1,1){20}}
\put(195,50){\line(-1,1){20}}
\put(215,50){\line(-1,1){20}}
\put(35,70){\line(-1,1){20}} 
\put(55,70){\line(-1,1){20}}
\put(75,70){\line(-1,1){20}}
\put(95,70){\line(-1,1){20}}
\put(115,70){\line(-1,1){20}}
\put(135,70){\line(-1,1){20}}
\put(155,70){\line(-1,1){20}}
\put(175,70){\line(-1,1){20}}
\put(195,70){\line(-1,1){20}}
\put(215,70){\line(-1,1){20}}
\put(14,2){\scriptsize{1}}
\put(34,2){\scriptsize{2}}
\put(54,2){\scriptsize{3}}
\put(74,2){\scriptsize{1}}
\put(94,2){\scriptsize{2}}
\put(114,2){\scriptsize{3}}
\put(134,2){\scriptsize{1}}
\put(154,2){\scriptsize{2}} 
\put(174,2){\scriptsize{3}}
\put(194,2){\scriptsize{1}}
\put(214,2){\scriptsize{2}} 
\put(15.5,34){\scriptsize{4}}
\put(35.75,34){\scriptsize{5}}
\put(56.5,34){\scriptsize{6}}
\put(75.5,34){\scriptsize{4}}
\put(95.5,34){\scriptsize{5}}
\put(115.5,34){\scriptsize{6}}
\put(135.5,34){\scriptsize{4}}
\put(155.5,34){\scriptsize{5}}
\put(175.5,34){\scriptsize{6}}
\put(195.5,34){\scriptsize{4}}
\put(215.5,34){\scriptsize{5}}
\put(15.5,54){\scriptsize{7}}
\put(35.75,54){\scriptsize{8}}
\put(56.5,54){\scriptsize{9}}
\put(75.5,54){\scriptsize{7}}
\put(95.5,54){\scriptsize{8}}
\put(115.5,54){\scriptsize{9}}
\put(135.5,54){\scriptsize{7}}
\put(155.5,54){\scriptsize{8}}
\put(175.5,54){\scriptsize{9}}
\put(195.5,54){\scriptsize{7}}
\put(215.5,54){\scriptsize{8}}
\put(15.5,74){\scriptsize{1}}
\put(35.75,74){\scriptsize{2}}
\put(56.5,74){\scriptsize{3}}
\put(75.5,74){\scriptsize{1}}
\put(95.5,74){\scriptsize{2}}
\put(115.5,74){\scriptsize{3}}
\put(135.5,74){\scriptsize{1}}
\put(155.5,74){\scriptsize{2}}
\put(175.5,74){\scriptsize{3}}
\put(195.5,74){\scriptsize{1}}
\put(215.5,74){\scriptsize{2}}
\put(14,94){\scriptsize{4}}
\put(34,94){\scriptsize{5}}
\put(54,94){\scriptsize{6}}
\put(74,94){\scriptsize{4}}
\put(94,94){\scriptsize{5}}
\put(114,94){\scriptsize{6}}
\put(134,94){\scriptsize{4}}
\put(154,94){\scriptsize{5}}
\put(175.5,94){\scriptsize{6}}
\put(195.5,94){\scriptsize{4}}
\put(215.5,94){\scriptsize{5}}
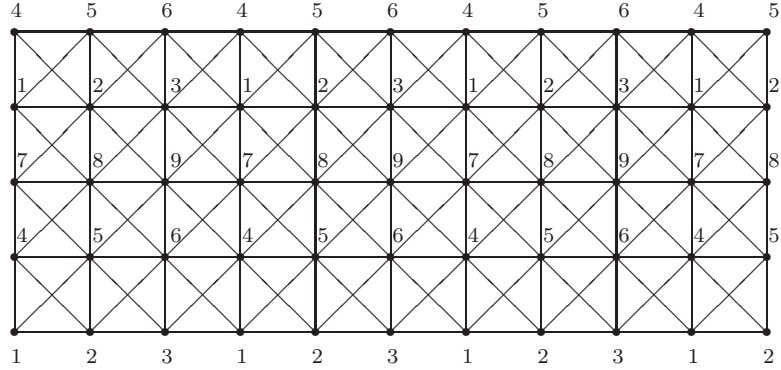
\put(28,-10){\footnotesize{\textbf{Fig.3.}} \footnotesize{Conditional $(9,8)$-coloring of $P_5 \otimes P_{11}$}}
\end{picture}
\end{center}
\bigskip
\subsection{Dynamic chromatic number of the $(t,n)$-web graph}
For $t \geq 1$ and $n \geq 3$, by $W(t,n)$ we denote the graph consisting of $t$ induced cycles $C_n$ such that no two $C_n$'s have a vertex in common and is constructed recursively as follows. Let $W(1,n)=W_{n+1}$, the wheel graph on $n+1$ vertices. We denote by $v_{0,0}$ the vertex of $W(1,n)$ with degree $n$. Assume that $W(t,n)$ has been obtained. Let $v_{t,1},v_{t,2},\ldots,v_{t,n}$ be the induced cycle in $W(t,n)$ such that for all $1 \leq i \leq n,\; d(v_{t,i})=3$ and  $v_{t,i}$ is adjacent to $v_{t,j}$ only if $|i-j|=1 \; \text{or} \; n-1$. We construct $W(t+1,n)$ from $W(t,n)$ with the following vertex and edge sets: 
\begin{align*}
V(W(t+1,n)) & =V(W(t,n)) \cup \{v_{t+1,1},v_{t+1,2},\ldots,v_{t+1,n}\}\; \; \text{and} \\
E(W(t+1,n)) & =E(W(t,n)) \cup \{(v_{t,i},v_{t+1,i}): 1 \leq i \leq n \} \cup E',
\end{align*}
where $E'=\{(v_{t+1,i},v_{t+1,j}): 1 \leq i,j \leq n \; \text{and} \; j-i=1 \; \text{or} \; n-1\}$.

\newtheorem{thm3}[thm1]{Theorem}
\begin{thm3}
Let $W(t,n)$ be a $(t,n)$- web graph. Then $\chi_d(W(t,n))=4$.
\end{thm3}  
\begin{proof}
For all $1 \leq i \leq n$, We denote the induced cycle $v_{i,1},v_{i,2},\ldots,v_{i,n}$ by $C_{i,n}$. Let $k=\chi_d(W(t,n))$. First we show that $k \geq 4$ by distinguishing the following two cases. \\
\textbf{Case} $1$ : $n:\;$odd. We know that if $H$ is a subgraph of $G$, then $\chi(H) \leq \chi(G) \leq \chi_r(G)$. Since $W_{n+1}$ is a subgraph of $W(t,n)$, taking $H=W_{n+1},G=W(t,n)$ and $r=2$ we have $\chi_d(W(t,n)) \geq \chi(W_{n+1})=4$.\\
\textbf{Case} $2$ : $n:\;$even. We assume the contradiction and let $k=3$. Let $c \colon V(W(t,n)) \to \{1,2,3 \}$ be a possible conditional $(3,2)-$coloring of $W(t,n)$. We consider possible colorings to a vertex such that always (C1) holds. W.l.o.g., we may assume that $c(v_{1,i})=1$ if $i$ is odd and $2$ otherwise. Then it forces that $c(v_{0,0})=3$ and $c(v_{2,2}) \neq 2$. We now distinguish two subcases: \\
\textbf{Case} $2.1$ : $c(v_{2,2}) = 3$. We must have $c(v_{2,1}) = 2$ and $c(v_{2,3}) = 2$. If $t=2$ then clearly  $v_{2,2}$ violates (C2). If $t>2$, (C2) is satisfied at $v_{2,2}$ only if $c(v_{3,2}) = 1$. This implies that $c(v_{3,1})=c(v_{3,3})=3$. Now if $t=3$ then (C2) is violated at $v_{3,2}$. Ingeneral, supposing that (C1) holds (C2) will be violated at $v_{t,2}$.\\
\textbf{Case} $2.2$ : $c(v_{2,2}) = 1$. Clearly $c(v_{2,1}) \neq 1$. Let $c(v_{2,1}) = 2$. If $t=2$ then we must have $c(v_{2,n}) = 3$ otherwise (C2) will be violated at $v_{2,1}$. This forces that $c(v_{2,n-1}) = 2$. Hence (C2) is violated at $v_{2,n}$. If $t>2$, then $c(v_{2,n}) = 1 \; \text{or} \; 3$. Suppose that $c(v_{2,n}) = 3$, this forces that $c(v_{2,n-1}) = 2$. If $t=3$ then (C2) is satisfied at $v_{2,n}$ only if  $c(v_{3,n}) = 1$. This forces that $c(v_{3,1}) = c(v_{3,n-1}) =3$ and implies a violation of (C2) at $v_{3,n}$. Ingeneral if $c(v_{2,n}) = 3$ then (C2) is violated at $v_{t,n}$. Similarly it can be easily shown that if $c(v_{2,n}) = 1$ then (C2) will be violated at $v_{t,1}$. Similarly it can be shown that if even when $c(v_{2,1}) = 3$, there exists a vertex at which (C2) is violated.\\
Hence from the cases $1$ and $2$ (above) it is clear that $k \geq 4$. \\ To show that $k \leq 4$ it suffices to construct a conditional $(4,2)-$coloring of $W(t,n)$. We now distinguishing four cases as given below. In all these cases we assign color $4$ to the vertex $v_{0,0}$ and assume that for all $1 \leq i \leq t$ the vertices of $C_{i,n}$ are colored in the order $v_{i,1},v_{i,2},\ldots,v_{i,n}$.\\
\textbf{Case} $1$: $n \equiv 0 \pmod{4}$. Color the vertices of $C_{1,n}$ by repeating the sequence of colors $1,2,1$ and $3$. The remaining uncolored vertices are colored as follows. If $i$ is even then the vertices of $C_{i,n}$ are colored  $v_{i,1}$ by repeating the sequence of colors $3,1,4$ and $2$. If $i$ is odd then the vertices of $C_{i,n}$ are colored by repeating the sequence of colors $1,4,2$ and $3$. \\
\textbf{Case} $2$: $n \equiv 1 \pmod{4}$. Color the first $4k$ vertices of $C_{1,n}$ by repeating the sequence of colors $1$ and $2$. There remains one uncolored vertex of $C_{1,n}$, to which we assign color $3$. If $i$ is even then the first $4k$ vertices of $C_{i,n}$ are colored by repeating the sequence of colors $3,1,2$ and $4$. The remaining vertex is assigned color $1$. If $i$ is odd then the first $4k$ vertices of $C_{i,n}$ are colored by repeating the sequence of colors $4,2,3$ and $1$. The remaining vertex is assigned color $3$.\\
\textbf{Case} $3$: $n \equiv 2 \pmod{4}$. Color the first $4k$ vertices of $C_{1,n}$ by repeating the sequence of colors $1$ and $2$. 
The remaining two vertices are assigned colors $1$ and $3$ in order. If $i$ is even then the first $4k$ vertices of $C_{i,n}$ are colored by repeating the sequence of colors $3,4,2$ and $1$. The remaining two vertices are assigned colors $4$ and $2$ in order. If $i$ is odd then the first $4k$ vertices of $C_{i,n}$ are colored by repeating the sequence of colors $1,2,3$ and $4$. The remaining two vertices are assigned colors $2$ and $3$ in order.\\
\textbf{Case} $4$: $n \equiv 3 \pmod{4}$. Color the first $4k+2$ vertices of $C_{1,n}$ by repeating the sequence of colors $1$ and $2$. 
The remaining vertex is assigned color $3$. If $i$ is even then the vertices of $C_{i,n}$ are colored by repeating the sequence of colors $2,3,4$ and $1$. If $i$ is odd then the first $4k$ vertices of $C_{i,n}$ are colored by repeating the sequence of colors $3,1,2$ and $4$. The remaining three vertices are assigned colors $3,2$ and $1$ in order.\\
It can be checked that the coloring assignment given in the four cases above is a conditional $(4,2)$-coloring of $W(t,n)$. Therefore $k \leq 4$, and altogether we have $k=4$. Hence the result.
\end{proof}

\end{document}